\documentclass{elsart}

\usepackage{cite}
\usepackage{graphicx}

\begin{document}

\begin{frontmatter}

\title{Homotopy perturbation method: when infinity equals five}

\author{Francisco M. Fern\'{a}ndez \thanksref{FMF}}

\address{INIFTA (UNLP,CCT La Plata-CONICET), Divisi\'{o}n Qu\'{i}mica Te\'{o}rica,\\
Diag. 113 y 64 (S/N), Sucursal 4, Casilla de Correo 16,\\
1900 La Plata, Argentina}

\thanks[FMF]{e--mail: fernande@quimica.unlp.edu.ar}

\begin{abstract}
I discuss a recent application of homotopy perturbation method
to a heat transfer problem. I show that the authors make infinity
equal five and analyze the consequences of that magic.
\end{abstract}

\end{frontmatter}

There has recently been great interest in the application of several
approximate procedures, like the homotopy perturbation method (HPM), the
Adomian decomposition method (ADM), and the variation iteration method
(VIM), to a variety of linear and nonlinear problems of interest in
theoretical physics \cite
{RDGP07,CHA07,YO07,CH07a,EG07,GAHT07,SNH07,CH08,ZLL08,SNH08b,M08,SNH08,RAH08,YE08,SG08}%
. From now on I will refer to those variation and perturbation approaches as
VAPA. In a series of papers I have shown that most of the VAPA\ results are
useless, nonsensical, and worthless\cite{F07,F08b,F08c,F08d,F08e}. In many
of those papers the authors try to solve nonlinear problems by means of
elaborated VAPA implementations that merely produce the Taylor expansion of
the solutions. Of course, such approximate expressions do not give the
overall picture of the dynamics, and the authors are merely content with a
description of the initial stages of the evolution which do not tell us
anything relevant about the process. Other authors solve the Schr\"{o}dinger
equation and obtain trivial unphysical solutions that are not square
integrable. As an example I mention two great feats of the VAPA users: the
expansion of exponential functions of the form $e^{i\alpha t}$\cite{SG08}
and a prey--predator model that predicts a negative population of rabbits%
\cite{YE08} (see also my comments\cite{F08c,F08d}).

However, my criticisms have not been welcome because they lack ``the
qualities of significant timeliness and novelty that we are seeking in this
journal'' and for that reason they remain unpublished outside arXiv.

The purpose of this article is the analysis of a recently published paper
that certainly meets the criterion of timeliness and novelty sought in that
journal. Esmaeilpour and Ganji\cite{EG07} applied homotopy perturbation
method (HPM) to the solution of the problem of forced heat convection over
an horizontal flat plate. After some algebraic manipulation of the
Navier--Stokes equations they obtained two coupled nonlinear differential
equations:\cite{EG07}
\begin{eqnarray}
f^{\prime \prime \prime }(\eta )+\frac{1}{2}f(\eta )f^{\prime \prime }(\eta
) &=&0  \nonumber \\
\varepsilon \theta ^{\prime \prime }(\eta )+\frac{1}{2}f(\eta )\theta
^{\prime }(\eta ) &=&0  \label{eq:diffeq}
\end{eqnarray}
with the boundary conditions
\begin{eqnarray}
f(0) &=&f^{\prime }(0)=0,\,f^{\prime }(\infty )=1  \nonumber \\
\theta (0) &=&1,\,\theta (\infty )=0  \label{eq:BC}
\end{eqnarray}

The HPM yields series of the form
\begin{equation}
f=\sum_{j=0}^{\infty }f_{j}p^{j},\,\theta =\sum_{j=0}^{\infty }\theta
_{j}p^{j}  \label{eq:HPM_series}
\end{equation}
where the dummy perturbation parameter $p$ is set equal to unity at the end
of the calculation. Esmaeilpour and Ganji\cite{EG07} choose the boundary
conditions
\begin{eqnarray}
f_{j}(0) &=&0,\,f_{j}^{\prime }(0)=0,\,f_{j}^{\prime }(\infty )=\delta _{j0}
\nonumber \\
\theta _{j}(0) &=&\delta _{j0},\,\theta _{j}(\infty )=0  \label{eq:HPM_BC}
\end{eqnarray}

Surprisingly, the perturbation corrections $f_{j}(\eta )$ and $\theta
_{j}(\eta )$ are polynomial functions of $\eta $\cite{EG07} which cannot
satisfy the boundary conditions at infinity (\ref{eq:HPM_BC}) although the
authors appear to state otherwise\cite{EG07}. In fact, the approximate
function
\begin{equation}
f_{^{HPM}}(\eta )=\frac{1348969}{7741440}\eta ^{2}-\frac{4867}{10752000}\eta
^{5}+\frac{451}{322560000}\eta ^{8}-\frac{1}{532224000}\eta ^{11}
\label{eq:HPM_f}
\end{equation}
corrected to third order ($j\leq 3$) does not satisfy the boundary
conditions (\ref{eq:BC}). However, the figures shown by Esmaeilpour and Ganji%
\cite{EG07} exhibit a reasonable agreement between the exact and approximate
solutions for $0\leq \eta \leq 5$.

\begin{figure}[H]
\begin{center}
\includegraphics[width=9cm]{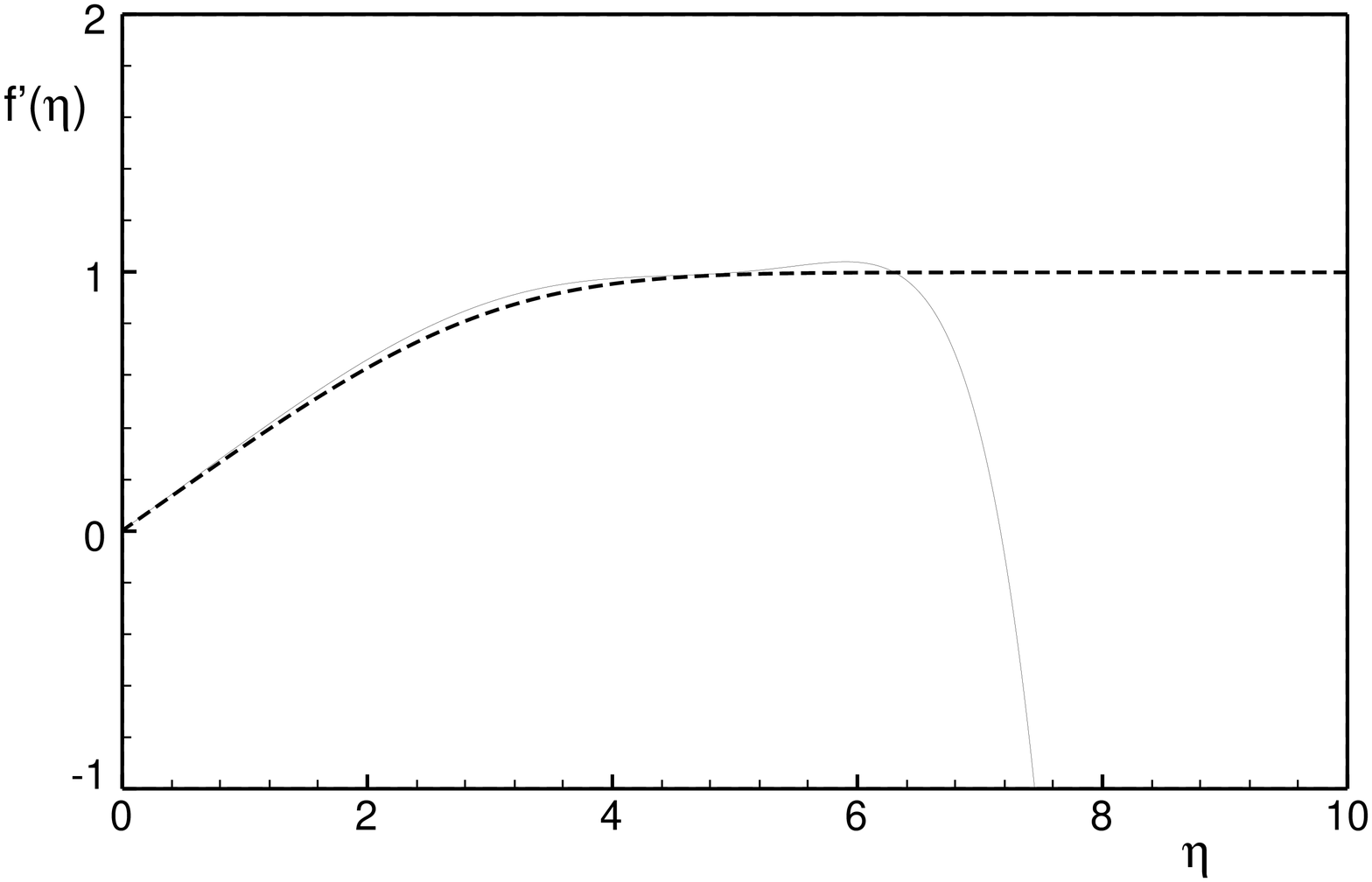}
\end{center}
\caption{Numerical (dashed line) and HPM (solid line) values of $%
f^{\prime}(\eta)$ }
\label{fig:EG1}
\end{figure}

When VAPA does not fit the problem the users make the problem fit VAPA. In
this case Esmaeilpour and Ganji\cite{EG07} do some kind of magic and make
infinity equal five. Consequently, their approximate solutions satisfy the
following boundary conditions
\begin{eqnarray}
f_{j}(0) &=&0,\,f_{j}^{\prime }(0)=0,\,f_{j}^{\prime }(5)=\delta _{j0}
\nonumber \\
\theta _{j}(0) &=&\delta _{j0},\,\theta _{j}(5)=0  \label{eq:HPM_BC_5}
\end{eqnarray}
Unfortunately, the authors forgot to say how they did this miracle. Since I
am not that smart and still think that there is something else beyond that
shrunk infinity I produced Fig.~\ref{fig:EG1} that shows the actual
behaviour of $f_{HPM}^{\prime }(\eta )$ in a wider interval.

When solving the differential equation for $f$ one has to determine the
value of $f^{\prime \prime }(0)$ that is consistent with the boundary
condition at infinity. Esmaeilpour and Ganji\cite{EG07} do not discuss the
calculation of this unknown parameter although they obtained the numerical
solution by a standard software. Our straightforward approximate calculation
based on trial and error suggests that $f^{\prime \prime }(0)\approx
0.3320574$ and the HPM function (\ref{eq:HPM_f}) yields $f_{^{HPM}}^{\prime
\prime }(0)=0.349$. The discrepancy is probably due to the fact that I have
not been able to enter the shrunk infinity discovered by the authors. I
suppose that for this very reason  my contribution cannot be considered to
carry the qualities of significant timeliness and novelty.

It is my opinion that VAPA have produced one of the greatest concentrations
of bad papers I have ever seen. If the reader proves me wrong I will
certainly apologize.

\end{document}